\pdfoutput=1
\documentclass[a4paper,11pt]{article}

\newcommand\papertitle{\boldmath 
	Charged Moments in $W_3$ Higher Spin Holography
}

\usepackage{jheppub}
\usepackage{xcolor}
\usepackage[english]{babel}
\usepackage{textcomp}
\usepackage{amsmath,amsfonts,amssymb,mathrsfs}
\usepackage{bbm}
\usepackage[inline]{enumitem}
\usepackage{multirow}
\usepackage{subcaption}
\usepackage{nicefrac}
\usepackage{xspace}
\usepackage{mathtools}
\usepackage{graphicx} 
\usepackage{esint}
\graphicspath{{figures/}} 

\usepackage[colorlinks=true]{hyperref}
\hypersetup{
	bookmarks=true,         
	unicode=false,          
	pdftoolbar=true,        
	pdfmenubar=true,        
	pdffitwindow=false,     
	pdfstartview={FitH},    
	pdftitle={\papertitle}, 
	pdfauthor={Christian Northe},
	pdfnewwindow=true,      
	colorlinks=true,        
	linkcolor=blue,         
	citecolor=red,          
	filecolor=magenta,      
	urlcolor=cyan           
}

\usepackage[colorinlistoftodos,prependcaption,textsize=tiny]{todonotes}

\makeatletter

\@addtoreset{equation}{section}
\makeatother

\newcommand{\R}	{\mathbb{R}}

\newcommand{\cA}{\mathcal{A}}

\newcommand{\cH}{\mathcal{H}}

\newcommand{\cW}{\mathcal{W}}

\DeclareMathOperator{\Tr}{Tr}
\DeclareMathOperator{\tr}{\tr}

\newcommand{\ket}[1]	{\vert{#1}\rangle}

\renewcommand{\Tr}	{\mathrm{Tr}}
\renewcommand{\tr}	{\mathrm{tr}}

\newcommand\id		{\mathbf{1}}

\newcommand\Algebra[1]	{\mathfrak{#1}}

\newcommand{\SLNR}	{SL(N,\mathbb{R})}

\newcommand{\su}	{\Algebra{su}}

\newcommand\ads {\text{AdS}}

\newcommand{\AdS}	{\ads}

\newcommand{\kcs}	{k_{cs}}

\newcommand{\bbra}[1]	{\langle\!\langle{#1}\lVert}
\newcommand{\bket}[1]	{\lVert{#1}\rangle\!\rangle}

\newcommand{\cc}        {\text{c.c}}

\newcommand\secref[1]	{section~\ref{#1}\xspace}

\newcommand\appref[1] {appendix~\ref{#1}\xspace}

\begin{document}
	\title{\papertitle}

	\author[a,1,]{Suting Zhao,\note[1]{suting.zhao@physik.uni-wuerzburg.de}}
	\author[a,2,]{Christian Northe,\note[2]{christian.northe@physik.uni-wuerzburg.de}}
	\author[a,3]{Konstantin Weisenberger,\note[3]{konstantin.weisenberger@physik.uni-wuerzburg.de}}
	\author[a,4]{Ren\'e Meyer,\note[4]{Corresponding author: rene.meyer@physik.uni-wuerzburg.de
			\\
			The ordering of authors is chosen to reflect their role in the preparation of this work.}\\
	}
	\affiliation[a]{Institut f\"ur Theoretische Physik und Astrophysik\\ and \\
		W\"urzburg-Dresden Cluster of Excellence ct.qmat,\\ Julius-Maximilians-Universit\"at W\"urzburg, \\Am Hubland \\97074 W\"urzburg, Germany}
	
	\abstract{
		We consider the charged moments in $SL(3,\mathbb{R})$ higher spin holography, as well as in the dual two-dimensional conformal field theory with $W_3$ symmetry. For the vacuum state and a single entangling interval, we show that the $W_3$ algebra of the conformal field theory induces an entanglement $W_3$ algebra acting on the quantum state in the entangling interval. The algebra contains a spin 3 modular charge which commutes with the modular Hamiltonian. The reduced density matrix is characterized by the modular energy and modular charge, hence our definition of the charged moments is also with respect to these conserved quantities. We evaluate the logarithm of the charged moments perturbatively in the spin 3 modular chemical potential, by computing the corresponding connected correlation functions of the modular charge operator up to quartic order in the chemical potential. This method provides access to the charged moments without using charged twist fields.	Our result matches known results for the charged moment obtained from the charged topological black hole picture in  $SL(3,\mathbb{R})$ higher spin gravity. Since our charged moments are not Gaussian in the chemical potential any longer, we conclude that the dual $W_3$ conformal field theories must feature breakdown of equipartition of entanglement to leading order in the large $c$ expansion. 
	}
	
	
	\maketitle
	\flushbottom
	
	\section{Introduction}\label{sec:Intro}
	
	In recent years \cite{VanRaamsdonk} it has become clear that the quantum information content of the dual QFT state is an important key in reconstructing the bulk geometry in AdS/CFT. A prominent role is played by the entanglement entropy, whose AdS/CFT dual is given by a minimal area surface anchored at the boundary of the entangling region \cite{RT}, which can be used to reconstruct a large part (but not all \cite{Balasubramanian:2014sra, Abt:2018ywl}) of the bulk space-time. To gain further understanding of the relation between the quantum information structure of states in the boundary QFT and the bulk, it is useful to consider other quantum information measures, for a review c.f.~\cite{RangamaniTakayanagiBook}. One such measure is the symmetry-resolved entanglement entropy \cite{GoldsteinSela, Murciano:2020vgh, Murciano:2020lqq, Bonsignori:2020laa, Horvath:2020vzs,  tan2020particle, Capizzi:2020jed, Capizzi:2021kys, calabrese2021symmetry,Oblak:2021nbj,estienne2021finite, zhao2021symmetry,weisenberger2021symmetry}. It quantifies the entanglement in each subregion charge sector, providing a finer resolution of the information provided by the ordinary entanglement entropy. 
	
	The main tool for finding such a resolution is the charged moments, which was originally proposed in \cite{Belin:2013uta} as a charged generalization of the usual R\'enyi entropy. When restricting on the vacuum background with a single entangling interval, one can map the charged moments to a grand canonical thermal partition function in the boundary CFT, with its gravity dual given by the charged topological hole. It was further pointed out in \cite{zhao2021symmetry} that, instead of mapping to the thermal system, inserting a bulk $U(1)$ Wilson line defects also captures the influence of the charged moments. Such a holographic construction does not require the existence of a $U(1)$ isometry along the thermal cycle, but only relies on the replica symmetry for the bulk field configurations. This feature allows to construct the holographic dual of the charged moments in excited state backgrounds, as well as to analyse the phase transition in the charged moments for multiple entangling intervals from the holographic point of view. In all cases considered, \cite{zhao2021symmetry,weisenberger2021symmetry} found agreement of the calculated symmetry-resolved entanglement between the AdS and the CFT side of the correspondence. 
	
	One essential feature of the charged moments associated with $U(1)$ symmetries is that they are always Gaussian in the chemical potential \cite{zhao2021symmetry,weisenberger2021symmetry}. As a consequence, the resulting symmetry-resolved entanglement does not depend on the $U(1)$ charge. This charge independence is called equipartition of entanglement, implying that entanglement is equally distributed between all the charge sectors. This equipartition behavior was further discussed for WZW models with general compact group $G$ in \cite{calabrese2021symmetry}, where it was shown that equipartition of entanglement holds at leading order in the large interval expansion, but breaks down at the subleading constant order. Equipartition was also shown to break down in a one-dimensional fermionic chain with an arbitrary localized scattering region  \cite{fraenkel2021entanglement}. Currently, it is unclear which theories admit equipartition or not, and which properties of the theory are finally responsible for this. In this work, we shed some light on this question by finding hints towards a violation of equipartition of entanglement to leading order in the large $c$ limit in $SL(3,\mathbbm{R})$ higher spin holography.
	
	Higher spin theories have a long history in theoretical physics, for instance in the quantum Hall effect \cite{cappelli1994classification}, or in statistical systems such as the 3-states Potts model at criticality \cite{zamolodchikov1988integrals}. Furthermore, higher spin symmetry appears in toys models of classical and quantum gravity \cite{bergshoeff1990w, bouwknegt1993w, hull1991classical, schoutens1991induced}. While it is possible to construct theories of two-dimensional quantum gravity in which the higher spin algebra $\cW$ replaces the Virasoro algebra, termed $\cW$-gravity, higher spin fields naturally emerge in  the tensionless limit of string theory \cite{chang2013abj, gaberdiel2014higher,gaberdiel2016string}. Being a part of string theory, it is perhaps no surprise that higher spin symmetry also arises in bottom-up examples of gauge/gravity duality \cite{Maldacena}. Most prominently, there exists a duality between the large $N$ limit of $\cW_N$ minimal models,
	\begin{equation}
		\frac{\hat{\su}(N)_k\oplus\hat{\su}(N)_1}{\hat{\su}(N)_{k+1}}\,,
		\quad
		\text{where } k,\,N\to\infty\quad \text{such that}
		\quad
		\lambda=\frac{N}{k+N}=\text{const.}\in[0,1],\notag
	\end{equation}
	and three-dimensional Vasiliev theories \cite{vasiliev2003nonlinear} with gauge algebra\footnote{The algebra $\mathfrak{hs}[\lambda]$ is a generalization of $sl(N,\mathbb{R})$ from $N\in\mathbb{N}$ to $N\in\R$. The large $N$ limit of the coset is also referred to as $\cW_\infty$.} $\mathfrak{hs}[\lambda]$\cite{gaberdiel2011ads, gaberdiel2013minimal, gaberdiel2012triality}. Due to the infinite number of conserved charges implied by integrability, the CFT side is under good control, providing insight into the $\mathfrak{hs}[\lambda]$ theories via the duality. The latter can be viewed as toy models of quantum gravity, which are simpler than full-fledged string theory. Intriguingly, these dualities do not rely on supersymmetry, hinting that it is not a vital ingredient of the AdS/CFT correspondence. Hence, higher spin symmetry opens a window into new, yet tractable, regimes of the AdS/CFT correspondence, perhaps allowing to understand better the mechanisms underlying the duality. 
	
	A natural starting point to reach this duality is regular Chern-Simons gravity based on the gauge group $\SLNR \times \SLNR$ \cite{blencowe1989consistent, bergshoeff1990area, campoleoni2010asymptotic}. For $N=2$ one obtains standard  three-dimensional gravity with negative cosmological constant, where after fixing Brown-Henneaux boundary conditions, the asymptotic symmetry algebra is given by the Virasoro algebra \cite{Brown:1986nw, Grumiller:2016pqb}.  For $N \geq 3$, the theory contains massless interacting fields with spins ranging between $2$ and $N$ \cite{campoleoni2010asymptotic, campoleoni2011asymptotic}, constituting the gauge sector of more general higher spin theories. While evidence has been found that for finite $N>2$ these theories and their holographic dual CFTs show acausalities \cite{Perlmutter:2016pkf} unless embedded in larger higher spin theories including all spins, or directly string theory, it is nevertheless interesting to study them as subsectors of larger higher spin theories or even a full string theory construction. In this paper, being interested in the quantum information properties of these higher spin theories, we focus on the simplest of these cases, $N=3$, in which only a single additional particle of spin 3 appears in the spectrum.
	
In this work, we consider the charged moments for a single entangling interval in $SL(3,\mathbbm{R})$ higher spin holography, and in the dual two-dimensional conformal field theory with $W_3$ symmetry. We consider the vacuum case, and follow \cite{Belin:2013uta} by conformally mapping the calculation of the charged moments to the problem of calculating the grand canonical partition function of a topological higher spin black hole in the $SL(3,\mathbbm{R})$ higher spin gravity theory. This partition function is known \cite{Kraus:2011ds} for small spin 3 chemical potential, while at larger chemical potentials, the system admits a phase transition to another spin 3 black hole phase with a slightly different symmetry, namely $W_3^{(2)}$ algebra \cite{David:2012iu}. Since the partition function is not known all the way from small to large chemical potentials, we restrict to work perturbatively in small chemical potential. We use the $W_3$ algebra acting on the reduced density matrix on the entanglement interval to provide a novel CFT derivation of the charged moments. 
 
In this course, we define a modular charge operator on the entangling interval analogous to the definition of the modular Hamiltonian. This modular charge is invariant under modular flow and is the dual to the corresponding chemical potential in the charged moment. In particular, the modular charge is the zero mode of the $W_3$ conserved current on the subregion. Choosing maximally symmetric conformal boundary conditions then conserves the full symmetry algebra on the entangling interval and allows us to define the entanglement $W_3$ algebra on the subregion. We use the entanglement $W_3$ algebra to evaluate the charged moments by reducing them order by order in the chemical potential to connected correlation functions of modular charge operator insertions. While this reduction is possible in principle to an arbitrarily high order in the chemical potential, the integrals involved become technically more and more involved at higher orders. We calculate up to quartic order in the chemical potential expansion, and show that the corresponding charged moment matches with known results for the corresponding topological higher spin black hole. Our CFT calculation hence constitutes a new independent check of the AdS$_3$/CFT$_2$ correspondence for $SL(3,\mathbbm{R})$ higher spin gravity, which in principle also can be extended to higher orders in chemical potential.
	
	This paper is organized as follows: In section~\ref{sec:2} we review some salient features of higher spin gravity in \AdS$_{3}$. In section~\ref{sec:3} we first discuss the entanglement $W_3$ algebra, and comment on its relation to the BCFT picture of the entanglement entropy. We then discuss the calculation of the charged moments via the topological black hole in higher spin holography. In section~\ref{sec:4}, we use the entanglement $W_3$ algebra to derive the corresponding charged moment to quartic order in the chemical potential, which exactly matches the gravity result to that order in the chemical potential. We conclude in section~\ref{sec:conclusion}, and give an outlook to several future research directions. The appendices contain our conventions, detailed elaborations on the entanglement $W_3$ algebra and an alternative way of phrasing the Rényi entropies directly in terms of defect lines.
	
	\section{Higher spin theories in \AdS$_{3}$}\label{sec:2}
	Three-dimensional gravity with a negative cosmological constant can be formulated as a Chern-Simons theory with gauge group $SL(2,\mathbb{R})\times SL(2,\mathbb{R})$, c.f.  \cite{achucarro1986chern, witten19882+, witten2007three}. This formulation in terms of the Chern-Simons language is naturally extended to the higher spin case, i.e. $\SLNR \times \SLNR$ higher spin theory. This theory includes a tower of fields of spin $s\le N$ coupled to the graviton, and resembles the pure gravity case in many ways \cite{blencowe1989consistent}. The precise field content and the associated symmetry algebra of the dual CFT depends on how the $sl(2,\mathbb{R})$ subalgebra for the pure gravitational sector is embedded into $sl(N,\mathbb{R})$, as well as what kind of boundary conditions are imposed at spatial infinity. For the purpose of this paper, we focus on the $N=3$ case with the principle embedding of $sl(2,\mathbb{R})$ into $sl(3,\mathbb{R})$, for which the asymptotic symmetry algebra for the higher spin theory is a $W_3$ conformal algebra \cite{campoleoni2010asymptotic}. In this section, we briefly review some aspects of higher spin 3 gravity in Euclidean signature, that are relevant for this work.
	\subsection{$SL(N,\mathbb{R})\times SL(N,\mathbb{R})$ higher spin gravity}\label{sec:2.1}
	In the Euclidean formulation of the $\SLNR \times \SLNR$ Chern-Simons theory, the bulk action is given by
	\begin{align}
		I_{\text{bulk}}=i I_{CS}[A]-i I_{CS}[\bar{A}]\ ,
	\end{align}
	where 
	\begin{align}
		I_{CS}[A]=\frac{\kcs}{4\pi}\int_{\mathcal{M}}\Tr\left[A\wedge dA+\frac{2}{3}A\wedge A\wedge A\right]\ ,
	\end{align}
	and the connections $A$ and $\bar{A}$ are complex-valued 
	in $sl(N,\mathbb{R})$ \cite{Bunster:2014mua} in Euclidean signature with the relation $\bar{A}=-A^\dag$. Here, the trace is taken over the fundamental representation of $sl(N,\mathbb{R})$. The Chern-Simons level $\kcs$ is determined by matching the $sl(2,\mathbb{R})$ sector of the theory to pure Einstein-Hilbert gravity with $\AdS$ radius $l_{\AdS}$ and three-dimensional Newton constant $G_3$, yielding  
	\begin{align}
		\kcs=\frac{l_{\AdS}}{8G_3 \Tr[L_0 L_0]}\,.
	\end{align}
	Here $L_0$ is the Cartan element of the $sl(2,\mathbb{R})$ subalgebra in the fundamental representation of $sl(N,\mathbb{R})$. Hence, for fixed $\AdS$ radius, $\kcs$ depends on the choice of embedding.
	
	Furthermore, since the higher spin theory recovers pure Einstein-Hilbert gravity if all the higher spin fields are switched off, the central charge, as determined from the conformal transformation law of the boundary stress tensor, still is the standard Brown-Henneaux central charge 
	\begin{equation}
		c=\frac{3l_{\AdS}}{2G_3}\ .
	\end{equation}
	This leads to the relation
	\begin{align}
		c=12{\kcs} \Tr[L_0 L_0] \label{eq:central-charge}\ .
	\end{align}
	From now on, we set $l_{\AdS}=1$ for convenience.
	\subsection{The $N=3$ case}\label{sec:2.2}
	Now we focus on the $N=3$ case with the principle embedding of $sl(2,\mathbb{R})$. Following the conventions of \cite{hijano2014new} for the representation of the $sl(3,\mathbb{R})$ algebra, the generators read
	\begin{align}
		&	L_1=-\sqrt{2}
		\left(\begin{array}{ccc}
			0&0&0\\
			1&0&0\\
			0&1&0
		\end{array}\right)\ ,\quad
		L_0=
		\left(\begin{array}{ccc}
			1&0&0\\
			0&0&0\\
			0&0&-1
		\end{array}\right)\ ,\quad
		L_1=\sqrt{2}
		\left(\begin{array}{ccc}
			0&1&0\\
			0&0&1\\
			0&0&0
		\end{array}\right) \ ,\nonumber\\
		&   W_1=-\frac{1}{\sqrt{2}}
		\left(\begin{array}{ccc}
			0&0&0\\
			1&0&0\\
			0&-1&0
		\end{array}\right)\ ,\quad
		W_0=\frac{1}{3}
		\left(\begin{array}{ccc}
			1&0&0\\
			0&-2&0\\
			0&0&1
		\end{array}\right)\ ,\quad
		W_{-1}=\frac{1}{\sqrt{2}}
		\left(\begin{array}{ccc}
			0&1&0\\
			0&0&-1\\
			0&0&0
		\end{array}\right)\ ,\nonumber\\
		&   W_2=2
		\left(\begin{array}{ccc}
			0&0&0\\
			0&0&0\\
			1&0&0
		\end{array}\right)\ ,\quad
		W_{-2}=2
		\left(\begin{array}{ccc}
			0&0&1\\
			0&0&0\\
			0&0&0
		\end{array}\right)\ \,,
	\end{align}
	with the properties $L_n=(-1)^n L_{-n}^\dag$ and $W_n=(-1)^{n}W_{-n}^\dag$.
	The vielbein and the spin connection are given by
	\begin{align}
		e=\frac{A-\bar{A}}{2}\ , \quad \quad \omega=\frac{A+\bar{A}}{2},
	\end{align}
	and the corresponding metric and spin 3 field are constructed as
	\begin{align}
		g_{\mu\nu}=\frac{1}{\Tr[L_0 L_0]}\Tr[e_{\mu}e_{\nu}]=\frac{1}{2}\Tr[e_{\mu}e_{\nu}]\ ,\quad \phi_{\mu\nu\sigma}=\frac{1}{3!}\Tr[e_{(\mu}e_{\nu}e_{\sigma)}]\ .
	\end{align}
	The equations of motion for the Chern-Simons theory,
	\begin{align}
		dA+A\wedge A=0 \ , \quad \quad d\bar{A}+\bar{A}\wedge\bar{A}=0\ ,
	\end{align}
	enforce that the on-shell connections are locally flat. Combining this result with the relation $\bar{A}=-A^\dag$, one can write the general solutions in the following form,
	\begin{align}
		A=h^{-1}dh\ ,\quad \quad \bar{A}=-dh^\dag (h^\dag)^{-1}\ .
	\end{align}
	with $h$ being an $SL(3,\mathbb{C})$ matrix. 
	
	Let $\rho$ now be the radial coordinate of the manifold which diverges when approaching the boundary, and $(z,\bar{z})$ as the boundary complex coordinates. For the solutions to describe asymptotically $\AdS$ spaces, one usually imposes radial gauge, with connections of the form
	\begin{align}
		A=b^{-1}a b+b^{-1}db\ ,\quad \bar{A}=b \bar{a}b^{-1}+b db^{-1}\,, \quad b=e^{\rho L_0}\ .
	\end{align}\label{eq: A general}
	Furthermore, the connections $a$ and $\bar{a}$ take the highest weight form
	\begin{align}\label{eq: a general}
		& a=\left(L_1-\frac{\mathcal{L}(z)}{4 \kcs}L_{-1}+\frac{\mathcal{W}(z)}{4 \kcs}W_{-2}\right)dz\ ,\nonumber\\
		& \bar{a}=\left(L_{-1}-\frac{\bar{\mathcal{L}}(\bar{z})}{4 \kcs}L_{1}-\frac{\bar{\mathcal{W}}(\bar{z})}{4 \kcs}W_{2}\right)d\bar{z}\ ,
	\end{align}
	satisfying $\bar{a}=-a^\dag$ given the conditions $\bar{\mathcal{L}}(\bar{z})=\mathcal{L}(z)^*$ and $\bar{\mathcal{W}}(\bar{z})=\mathcal{W}(z)^*$. In the case where $\mathcal{W}(z)=0$, the solutions \eqref{eq: a general} are the Ba\~nados geometries \cite{banados1999three} for pure Einstein gravity.
	
	As shown in \cite{campoleoni2010asymptotic,Henneaux:2010xg}, the higher spin analog of the usual asymptotically $\AdS$ boundary condition is to require that the form of \eqref{eq: a general} does not change under the allowed gauge transformations.
	This requirement reduces the boundary gauge degrees of freedom, yielding the following variation for $\mathcal{L}$ and $\mathcal{W}$, 
	\begin{align}\label{eq: transformation law 1}
		&\delta_{\epsilon}\mathcal{L}=\frac{c}{12}\epsilon'''+2\epsilon' \mathcal{L}+\epsilon \mathcal{L}'\ ,\nonumber\\
		&\delta_{\epsilon}\mathcal{W}=3\epsilon'\mathcal{W}+\epsilon \mathcal{W}'\ ,\nonumber\\
		&\delta_{\eta}\mathcal{W}=\frac{c}{144}\eta^{(5)}+\frac{5}{6}\eta'''\mathcal{L}+\frac{1}{6}\eta \mathcal{L}'''+\frac{5}{4}\eta''\mathcal{L}+\frac{3}{4}\eta'\mathcal{L}''+\frac{16}{c}\eta'\mathcal{L}^2+\frac{16}{c}\eta(\mathcal{L}^2)' .
	\end{align}
	where $\epsilon=\epsilon(z)$ and $\eta=\eta(z)$ are the complex gauge parameter associated with the $sl(3,\mathbb{R})$ elements $L_1$ and $W_2$. Since \eqref{eq: transformation law 1} can be identified with the transformation law of the stress tensor and spin 3 current under the conformal and higher spin transformation in a $W_3$ CFT, one concludes that the asymptotic symmetry algebra is given by two copies of the $W_3$ algebra, with $\mathcal{L}(z)$ and $\mathcal{W}(z)$ being identified as the expectation values of the stress tensor $T(z)$ and spin 3 current $W(z)$ in the dual boundary state\footnote{In fact, when we fix the identification of $\mathcal{W}(z)$ as $\langle W(z)\rangle$, there will be two different ways to identify the antiholomorphic spin 3 current, i.e. $\mathcal{\bar{W}}(\bar{z})=\mathcal{W}(z)^*=\langle(\Omega\bar{W})(\bar{z})\rangle=\pm\langle\bar{W}(\bar{z})\rangle$, with $\Omega$ denoting the automorphisms for $W_3$ algebra.} The parameters $\epsilon(z)$ and $\eta(z)$ are associated with the conformal and spin 3 transformation in the dual boundary CFT. The explicit OPEs and algebras are given in \appref{appendix A}. For the other diagonal embedding, where the asymptotic symmetry algebra is the Polyakov-Bershadsky $W_3^{(2)}$ algebra, c.f.  \cite{Bershadsky:1990bg,Polyakov:1989dm}.
	
	\section{Entanglement $W_3$ algebra and higher spin charged moments}\label{sec:3}
	In this section, we consider a CFT with $W_3$ symmetry, study the charge statistics and define the associated charged moments for a single entangling interval $\mathcal{A}$. We start with an explicit construction of a single copy of the $W_3$ algebra acting on the entangling region. This construction is a direct generalization of the entanglement Virasoro algebra discussed in \cite{hu2020emergent}. Recall that the usual entanglement entropy and R\'enyi entropies in the vacuum background are captured by the spectrum of the modular Hamiltonian $\mathcal{H}_{\mathcal{A}}$, which is proportional to the Virasoro zero mode $L_0^{(\mathcal{A})}$ on the entangling interval $\mathcal{A}$ up to an additive constant \cite{CardyTonni}. The existence of the full $W_3$ algebra on the interval $\mathcal{A}$ allows us to define a spin 3 modular charge $\mathcal{Q}_{\mathcal{A}}$, which is associated with the $W_0^{(\mathcal{A})}$ mode, and is also conserved along the modular flow. The spin 3 modular charge is the natural object to define the higher spin charged moments as a charged generalization of the R\'enyi entropy.
	\subsection{Modular charge and higher spin charged moments}\label{sec: 3.1}
	The purpose of this subsection is the introduction of the spin 3 modular charge and the definition of higher spin charged moments. A natural framework for the construction of the modular charge, and in fact the full entanglement algebra, is the BCFT picture, which we adopt in the following. We only mention the results in the main text relegating the details to \appref{app: W3algebra}. 
	
	Considering a two-dimensional CFT, we denote by $\mathcal{A}$ a single entangling interval with its endpoints located at $z=z_1$ and $z=z_2$. The reduced density matrix $\rho_{\mathcal{A}}$ can be formally expressed as
	\begin{align}
		\rho_{\mathcal{A}}=e^{-2\pi \mathcal{H}_{\mathcal{A}}}\ ,
	\end{align}
	where $\mathcal{H}_{\mathcal{A}}$ is the so-called modular Hamiltonian. The modular Hamiltonian is in general non-local, which means that it can not be constructed via local operators. However, in the case where the background state is the vacuum state, $\mathcal{H}_{\mathcal{A}}$ can be constructed locally via the stress tensor as \cite{CardyTonni}
	\begin{align}\label{eq: modular H}
		\mathcal{H}_\cA=-\int_{z_2+\epsilon}^{z_1-\epsilon}\frac{dz}{2\pi i}\xi^z T_{zz}+\int_{\bar{z_2}+\bar{\epsilon}}^{\bar{z_1}-\bar{\epsilon}}\frac{d\bar{z}}{2\pi i}\xi^{\bar{z}}T_{\bar{z}\bar{z}}+C\ ,
	\end{align}
	where $\epsilon$ denotes the short distance cut-off around the endpoints of $\cA$ chosen such that local rotational invariance is respected. The constant $C$ in \eqref{eq: modular H} ensures $\tr \rho_\cA=1$.
	
	The modular Hamiltonian \eqref{eq: modular H} generates the diffeomorphism along the Euclidean boost vector field $\xi^a$, with components of the vector $\xi^a$ given by
	\begin{align}\label{eq: boost vector}
		\xi^z=\xi(z)=\frac{i(z-z_1)(z-z_2)}{(z_2-z_1)}\ ,\quad \xi^{\bar{z}}=\bar{\xi}(\bar{z})=-\frac{i(\bar{z}-\bar{z}_1)(\bar{z}-\bar{z}_2)}{(\bar{z}_2-\bar{z}_1)}\ .
	\end{align}
	By the conformal mapping from the complex plane to the annulus, i.e. $z\to w=\log\left(\frac{z-z_2}{z_1-z}\right)=x+i t_E$, the boost vector \eqref{eq: boost vector} simply becomes the Euclidean time vector on the annulus, i.e. $\xi^a=\left(\frac{\partial}{\partial t_{E}}\right)^a$. Due to the compactness of the time direction, the reduced density matrix $\rho_{\mathcal{A}}$ is then related to a thermal density matrix by a conformal mapping from the complex plane to the cylinder. As first pointed out in \cite{hu2020emergent}, it is possible to construct a full Virasoro algebra on a single entangling interval $\mathcal{A}$. Importantly, its Virasoro zero mode $L^{\mathcal{(A)}}_0$ is proportional to the modular Hamiltonian \eqref{eq: modular H} up to an additive constant \cite{CardyTonni},
	\begin{align}\label{eq: zero mode L0}
		L^{\mathcal{(A)}}_0=&\frac{l}{\pi^2}\int_{z_2+\epsilon}^{z_1-\epsilon}dz \frac{(z-z_1)(z-z_2)}{z_1-z_2}T(z) +\frac{l}{\pi^2}\int_{\bar{z_2}+\bar{\epsilon}}^{\bar{z_1}-\bar{\epsilon}}d\bar{z} \ \frac{(\bar{z}-\bar{z}_1)(\bar{z}-\bar{z}_2)}{\bar{z}_1-\bar{z}_2} \bar{T}(\bar{z})+\frac{c}{24}(1+\frac{4l^2}{\pi^2})\ ,
	\end{align}
	where
	\begin{align}\label{systemSize}
		l=\log\left|\frac{z_1-z_2}{\epsilon}\right|\ .
	\end{align}
	
	We now turn to the definition of the spin 3 modular charge. In order to arrive at the proper definition, it is essential to understand some basics of boundary conformal field theory (BCFT) and its relation to the calculation of entanglement entropy. First, in any QFT, a regularization scheme must be chosen when a spatial slice, on which the CFT state is defined, is partitioned into a region $\cA$ and its complement \cite{OhmoriTachikawa}. For our purposes, this is achieved by imposing conformal boundary conditions $a$ and $b$ respectively on either end of the entangling interval $\cA$. In the $z$-plane, the boundary conditions are implemented on the cut-off circles $|z-z_1|=|\epsilon|$ and $|z-z_2|=|\epsilon|$, which excise $z_1$ and $z_2$ respectively, while in $w$-coordinates, they limit the annulus of length \eqref{systemSize}.
	
	What remains after regularization is the familiar setup of a BCFT on a finite interval of length $l$ (c.f. \eqref{systemSize}) and boundary condition $a$ and $b$ at its end points. Denoting the Hilbert space of this system by $\cH_{ab}$, the  partition function $\mathcal{Z}_{n}$ is given by \cite{OhmoriTachikawa, CardyTonni}
	\begin{align}\label{eq: thermal Z 1}
		\mathcal{Z}_n=\Tr_{\mathcal{H}_{ab}}\left[q^{L_0^{(\mathcal{A})}-\frac{c}{24}}\right]\ , \qquad q=e^{2\pi i \tau}\ ,\quad \tau=\frac{in\pi}{2l}\ .
	\end{align}
	Observe that the system size \eqref{systemSize} and the Rényi index $n$ are encoded in $q$. By swapping (Euclidean) time and space by a modular $S$ transformation, the conformal boundary conditions $a$ and $b$ are   implemented by initial and final boundary states $\bket{a_n},\bket{b_n}$. Hence, the thermal partition function \eqref{eq: thermal Z 1} can be expressed a transition amplitude between said boundary states \cite{Cardy:1989ir},
	\begin{align}
		\mathcal{Z}_n=  \bbra{a_n} \tilde{q}^{\frac{1}{2}(L_0+\bar{L}_0-\frac{c}{12})}\bket{b_n}, \qquad \tilde{q}=e^{2\pi i \tilde{\tau}}, \quad \tilde{\tau}=-\frac{1}{\tau}\,.
	\end{align}\label{eq: transition apmplitude 1}
	In the limit $\epsilon\to 0$ and thus $l\to \infty$, $q\approx1$ and $\tilde{q}\ll1$, the thermal partition function can be expanded in $\tilde{q}$ as 
	\begin{align}
		\mathcal{Z}_n
		\approx
		\tilde{q}^{-\frac{c}{24}}\bbra{a_n} 0\rangle \langle 0\bket{b_n}+\dots
		=
		\tilde{q}^{-\frac{c}{24}}g_{a,n}g_{b,n}+\dots\ ,
	\end{align}
	and $g_{a,n}=\bbra{a_n} 0\rangle$ as well as $g_{b,n}=\langle 0\bket{b_n}$ are the g-factors, whose logarithm are the Affleck-Ludwig boundary entropies \cite{AffleckGFactor} of the boundary states. Here we assumed that only the lowest lying state in $\bket{a_n}$ and $\bket{b_n}$ is the vacuum $\ket{0}$ with weights $h=\bar{h}=0$, and survives the $l\to \infty$. In holographic CFTs, however, the boundary entropies can be neglected in \eqref{eq: transition apmplitude 1} due to the large $l$ limit.
	
	In CFTs with extended symmetries, maximally symmetric boundary conditions  which retain one full copy of the chiral algebra of the CFT without boundaries can be found \cite{Recknagel:2013uja}. For the case of interest in this paper,  this amounts to the two copies of the $W_3$ algebra being broken down to a single copy by chosing an entangling interval $\cA$ and imposing the maximally symmetric boundary conditions. This generalizes the construction in \cite{hu2020emergent} to the case where the entanglement algebra is $W_3$. The details of the construction of the entanglement $W_3$ algebra using the BCFT formalism are given in \appref{app: W3algebra}. Here we only show the expression for the zero mode of the spin 3 current,
	\begin{align}\label{eq: zero mode W0}
		W^{(\mathcal{A})}_0=\frac{2l^2}{i\pi^3}\int_{z_2+\epsilon}^{z_1-\epsilon}dz \frac{(z-z_1)^2(z-z_2)^2}{(z_1-z_2)^2}W(z) -\frac{2l^2}{i\pi^3}\int_{\bar{z_2}+\bar{\epsilon}}^{\bar{z_1}-\bar{\epsilon}}d\bar{z} \ \frac{(\bar{z}-\bar{z}_1)^2(\bar{z}-\bar{z}_2)^2}{(\bar{z}_1-\bar{z}_2)^2} (\Omega\bar{W})(\bar{z}) \ ,
	\end{align}
	Here, as usual in BCFT, we have allowed for an automorphism $\Omega$ of the $W_3$ algebra, which acts on the spin 3 current as $(\Omega \bar{W})(\bar{z})=\pm \bar{W}(\bar{z})$. This is analogous to the well-known Neumann and Dirichlet boundary conditions in the free boson theory. We discuss in the next subsection how $\Omega$ enters naturally in a holographic theory. 
	
	Since the zero mode $W^{(\mathcal{A})}_0$ commutes with $L^{(\mathcal{A})}_0$ and hence with $\mathcal{H}_{\mathcal{A}}$, we further can define the following spin 3 modular charge as an analog of the modular Hamiltonian \eqref{eq: modular H},
	\begin{align}\label{eq: modular Q}
		\mathcal{Q}_{\mathcal{A}}&=-\int_{z_2+\epsilon}^{z_1-\epsilon}\frac{dz}{2\pi i}\xi^z\xi^z W_{zzz}+\int_{\bar{z_2}+\bar{\epsilon}}^{\bar{z_1}-\bar{\epsilon}}\frac{d\bar{z}}{2\pi i}\xi^{\bar{z}}\xi^{\bar{z}}\left(\Omega\bar{W}_{\bar{z}\bar{z}\bar{z}}\right)
		=
		\frac{\pi^2}{4l^2} W^{(\mathcal{A})}_{0}\,.
	\end{align}
	In order to study the charge statistics on the region $\mathcal{A}$, we consider the following higher spin charged moments
	\begin{align}\label{eq: charged moments}
		\Tr_{\mathcal{H}_{ab}}\left[\rho_{\mathcal{A}}^{n}e^{2\pi i\alpha \mathcal{Q}_{\mathcal{A}}}\right]=\Tr_{\mathcal{H}_{a b}}\left[e^{-2\pi n \mathcal{H}_{\mathcal{A}}+2\pi i\alpha \mathcal{Q}_{\mathcal{A}} }\right]=\frac{\mathcal{Z}_{n}(\alpha)}{\mathcal{Z}_{1}(0)^n}\ ,
	\end{align}
	where $\mathcal{Z}_{n}(\alpha)$ is grand canonical thermal partition function, given by
	\begin{align}\label{eq: Zn alpha}
		\mathcal{Z}_n(\alpha)=\Tr_{\mathcal{H}_{ab}}\left[q^{L_{0}^{(\mathcal{A})}}y^{W^{(\mathcal{A})}_{0}}\right]\ ,
		\qquad
		y=e^{2\pi i \mu}, \quad \mu=\frac{\pi^2\alpha}{4l^2}\,.
	\end{align}
	In order to capture the $\alpha$-dependence of the charged moments $\mathcal{Z}_{n}(\alpha)$, it is useful to introduce the normalized generating function,
	\begin{align}\label{eq: generating function}
		f_{n}(\alpha):=\frac{\mathcal{Z}_{n}(\alpha)}{\mathcal{Z}_{n}(0)}\,,
	\end{align}
	with the initial condition $f_{n}(0)=1$. It is this generating function that we will compute on the gravity side in \secref{sec: 3.2}, and on the CFT side in \secref{sec:4}.
	

	\subsection{Holographic calculation}\label{sec: 3.2}
	Since in the cylinder coordinates $w=x+i t_E$ the charged moment \eqref{eq: charged moments} is, up to a normalization constant, nothing but the grand canonical thermal partition function \eqref{eq: Zn alpha}. The corresponding gravity dual is given by the topological higher spin 3 black hole.\footnote{In principle, the thermal partition function \eqref{eq: charged moments} also includes the boundary entropy contribution. However, this contribution will be subleading in the large interval length limit $\ell\rightarrow\infty$. Moreover, the boundary entropy contributions are scheme independent, and in particular do not depend on the R\'enyi index $n$ \cite{CardyTonni}. We hence neglect these boundary effects and only  focus on the leading $\log \frac{\ell}{\epsilon}$ term in the charged moment. Precise calculations for boundary states as well as boundary entropy require a full top-down description for the theory.} The same argument in the case of a $U(1)$ charge was first proposed in \cite{belin2013holographic}. The higher spin black hole solution on a solid torus was first discussed in \cite{Gutperle:2011kf,Kraus:2011ds}.
	
	Although the topology of the charged topological black hole is a solid cylinder, with spatial length $\Delta x=2l=2\log{|\frac{z_1-z_2}{\epsilon}|}$ and time period $\Delta t_{E}=2\pi n$, the compactness of the spatial direction does not have a strong influence on the thermodynamics in the limit $\Delta x\to \infty$. Note that there are two different formalisms for defining the higher spin black hole thermodynamics, the holomorphic formalism \cite{Gutperle:2011kf} and the canonical formalism \cite{Bunster:2014mua}. Here we work in the canonical formalism, in which the conserved charges and the first law of the black hole thermodynamics can be derived conveniently through the Iyer-Wald formalism \cite{hijano2014new,Iyer:1994ys}. In the cylinder coordinates $w=x+it_E$, the components of the connection $a$ are given by
	\begin{align}\label{topoBHsolution}
		& a_{x}=L_1-\frac{\mathcal{\mathcal{L}}}{4 \kcs}L_{-1}+\frac{\mathcal{W}}{4 \kcs}W_{-2}\ , &
		& a_{t_E}=i a_{x}+\frac{\alpha}{n}\left(a_{x}^2-\frac{1}{3}\Tr[a_{x}^2]\right)\ ,\nonumber\\
		& \bar{a}_{x}=L_{-1}-\frac{\bar{\mathcal{L}}}{4 \kcs}L_1-\frac{\bar{\mathcal{W}}}{4 \kcs}W_2\ , &		& \bar{a}_{t_E}=-i\bar{a}_{x}-\frac{\bar{\alpha}}{n}\left(\bar{a}_{x}^2-\frac{1}{3}\Tr[\bar{a}_{x}^2]\right)\ ,
	\end{align}
	where $\bar{\alpha}$ is the complex conjugate of $\alpha$. In order to show that the partition function of this solution \eqref{topoBHsolution} coincides with the saddle point approximation of the CFT thermal partition function \eqref{eq: Zn alpha}, we first need to identify the correct formulas for the black hole mass, spin 3 charge and entropy. Several different formalisms exist in the literature \cite{hijano2014new,Banados:2012ue,Campoleoni:2012hp,Perez:2012cf,deBoer:2013gz}. For consistencency of this paper, we present here the procedure in the Iyer-Wald formalism, following the original derivation in \cite{hijano2014new}.
	
	The starting point in the Iyer-Wald formalism is to consider the symplectic current for the Chern-Simons theory, which is obtained by a second variation of the Chern-Simons action, given by
	\begin{align}\label{eq:symplectic}
		\omega(\delta_1 A, \delta_{2} A)=-\frac{i k_{cs}}{2\pi}\Tr(\delta_1 A\wedge\delta_2 A)\,.
	\end{align}
	The variation $\delta_2$ in \eqref{eq:symplectic} is viewed as generated by an infinitesimal gauge transformation with gauge parameter $\Lambda$, i.e. $\delta_2=\delta_{\Lambda}$. This yields
	\begin{align}\label{eq: omega}
		\omega(\delta A, \delta_{\Lambda} A)=\frac{i k_{cs}}{2\pi}\Tr(d(\Lambda\delta A)-\Lambda\delta F)\,.
	\end{align}
	In the case that $\Lambda$ is a symmetry of the background solution, in the sense that $\delta_\Lambda A=0$, and the variation $\delta A$ is on-shell, i.e. $\delta F=0$, then one can integrate \eqref{eq: omega} over a constant time slice of the black hole with inner boundary at the bifurcation surface $\Sigma$ and outer boundary at spatial infinity. This yields
	\begin{align}\label{eq: omega 2}
		\frac{i k_{cs}}{2\pi}\int_{\infty}\Tr(\Lambda\delta A)=\frac{i k_{cs}}{2\pi}\int_{\Sigma}\Tr(\Lambda\delta A)\,.
	\end{align}
	In Wald's original derivation \cite{Iyer:1994ys}, the diffeomorphism which generates the variation $\delta_2$ is chosen to be time translation. The gauge parameter $\Lambda$ associated with the time translation is given by
	\begin{align}\label{eq: Lambda}
		\Lambda=A_{a}\left(\frac{\partial}{\partial_{t_E}}\right)^a=A_{t_E}=b^{-1}a_{t_{E}}b\,.
	\end{align}
	An analogous relation holds for the barred counterpart, $\bar{\Lambda}=\bar{A}_{t_E}$. 
	
	For the solution \eqref{topoBHsolution}, inserting \eqref{eq: Lambda} into \eqref{eq: omega 2} and including the barred piece gives rise to the following first law
	\begin{align}\label{eq: first law}
		\delta S=2\pi n \delta H-2\pi i \text{Re}(\alpha)\alpha \delta Q^{(+)}+2\pi \text{Im}(\alpha) \delta Q^{(-)}\ ,
	\end{align}
	where $(S, H, Q^{(\pm)})$ are identified with the entropy, mass and spin 3 charges, given by\footnote{In fact, $\delta S$ in \eqref{eq: first law} is equivalent to the variation of $S$ in \eqref{eq: charges and entropy} only if the holomony condition \eqref{holonomy condition 1} is imposed. (See \cite{hijano2014new} for details.) This is analogous to the condition of constant surface gravity for the black hole horizon as used in Wald's original derivation \cite{Iyer:1994ys}.}
	\begin{align}\label{eq: charges and entropy}
		&S=-i k_{cs} \int_{-l}^{l} dx\ \Tr[(n a_{t_E})a_{x}-(n \bar{a}_{t_E})\bar{a}_{x}]=(2n\mathcal{L}+2n\bar{\mathcal{L}}+3 i \alpha \mathcal{W}-3 i\bar{\alpha} \bar{\mathcal{W}})l\ , \nonumber\\
		&H=\frac{ k_{cs}}{2\pi}\int_{-l}^{l} dx \ \frac{1}{2}\Tr[a_{x}^2+\bar{a}_{x}^2]=\frac{(\mathcal{L}+\bar{\mathcal{L}})l}{\pi}\ ,\nonumber\\
		&Q^{(\pm)}=\frac{\kcs}{2\pi}\int_{-l}^{l} dx \ \frac{1}{3}\Tr[a_{x}^3\pm\bar{a}_{x}^3]=\frac{(\mathcal{W}\pm\bar{\mathcal{W}})l}{\pi}\ .
	\end{align}
	Combining \eqref{eq: first law} and \eqref{eq: charges and entropy} the partition function of the charged topological black hole is obtained,
	\begin{align}\label{eq:gravityZnnew}
		\mathcal{Z}_{n}(\alpha)=e^{S-2\pi n H+2\pi i \text{Re}(\alpha) Q^{(+)}-2\pi \text{Im}(\alpha)Q^{(-)}}\ .
	\end{align}
	Note that there are two spin-3 charges $Q^{(+)}$ and $Q^{(-)}$ in \eqref{eq:gravityZnnew}, coupled to $\text{Re}(\alpha)$ and $\text{Im}(\alpha)$. Matching \eqref{eq:gravityZnnew} with the saddle point approximation of the CFT thermal partition function \eqref{eq: Zn alpha} requires the parameter $\alpha$ to be either real (and $Q^{(+)}$ the charge) or purely imaginary (and $Q^{(-)}$ the charge). Under this requirement, the gravity partition can be written as
	\begin{align}\label{eq: gravity Zn}
		\mathcal{Z}_{n}(\alpha)=e^{S-2\pi n H+2\pi i \alpha Q^{(\pm)}}\,,
	\end{align}
	with the identification
	\begin{align}
		H=\frac{\pi}{2l}\left\langle L_{0}^{(\mathcal{A})}-\frac{c}{24}\right\rangle_{n, \alpha}\ , \quad Q^{(\pm)}=\frac{\pi^2}{4l^2}\left\langle W^{(\mathcal{A})}_{0}\right\rangle_{n, \alpha}\,.
	\end{align}
	One can also express the zero modes \eqref{eq: zero mode L0} and \eqref{eq: zero mode W0} in the cylinder $w$-coordinates, which results in the following identification for the currents,
	\begin{align}
		\mathcal{L}=\left\langle T(w)\right\rangle_{n, \alpha}\ ,\ \bar{\mathcal{L}}=\left\langle \bar T(\bar{w})\right\rangle_{n, \alpha}\ ,\ \mathcal{W}=\left\langle W(w)\right\rangle_{n, \alpha}\ ,\ \bar{\mathcal{W}}=\pm\left\langle(\Omega \bar{W})(\bar{w})\right\rangle_{n, \alpha}\ .
	\end{align}
	Coming back to the solution \eqref{topoBHsolution}, the values of $(\mathcal{L},\mathcal{W})$ are fixed by imposing the holonomy condition around the thermal circle \cite{Gutperle:2011kf}, 
	\begin{align}\label{holonomy condition 1}
		e^{\int dt_E \ a_{t_E}}=e^{2\pi n a_{t_E}}\cong e^{2\pi i L_0}\ ,
	\end{align}
	where `$\cong$' denotes equivalence up to conjugation by an element of $SL(3,\mathbb{R})$. This is the gauge invariant higher spin version of the smoothness condition for the horizon of the Euclidean higher spin black hole. Equivalent conditions for \eqref{holonomy condition 1} are given by \cite{Gutperle:2011kf}
	\begin{align}\label{holonomy condition 2}
		\det[a_{t_{E}}]=0\ ,\quad \Tr[a_{t_E}^2]=-\frac{1}{n^2}\Tr[L_0^2]\ .
	\end{align}
	Inserting the topological black hole solution \eqref{topoBHsolution} into \eqref{holonomy condition 2} yields algebraic equations
	\begin{align}\label{holonomy condition 3}
		& 0=-2\mathcal{L}^3\alpha^3+27 i \kcs \mathcal{L} \mathcal{W} n \alpha^2-18 \kcs\mathcal{L}^2 \alpha n^2+27 \kcs \mathcal{W}^2\alpha^3-27 i \kcs^2 \mathcal{W}n^3\ ,\nonumber\\
		& 0=\mathcal{L}^2\alpha^2 +3 \kcs^2+9 i \kcs \mathcal{W} \alpha n-3 \kcs \mathcal{L}n^2\,.
	\end{align}
	There are different branches of solutions for \eqref{holonomy condition 3}, which were discussed in \cite{David:2012iu,Chen:2012ba}. Here we only consider the BTZ-branch, i.e. the solution that recovers the topological BTZ black hole when $\alpha=\bar{\alpha}=0$. In order to compare with results from the perturbative CFT calculation in \secref{sec:4}, we consider the region where the dimensionless parameter satisfies $\alpha/n^2\ll 1$. The expectation values $\mathcal{L}$ and $\mathcal{W}$ are given by
	\begin{align}
		\mathcal{L}&=\frac{ \kcs}{n^2}\left[1-\frac{5}{3}\frac{\alpha^2}{n^4}+\frac{10}{3}\frac{\alpha^4}{n^8}+\frac{211}{27}\frac{\alpha^6}{n^{12}}+\cdots\right]\ ,\nonumber\\
		\mathcal{W}&=\frac{i \kcs}{n^3}\left[-\frac{2}{3}\frac{\alpha}{n^2}+\frac{40}{27}\frac{\alpha^3}{n^6}-\frac{34}{9}\frac{\alpha^5}{n^{10}}+\cdots\right]\,.
	\end{align}
	The resulting partition function is then given by 
	\begin{align}
		\log\mathcal{Z}_{n}(\alpha)=\frac{cl }{6n}\left[1-\frac{1}{3}\frac{\alpha^2}{n^4}+\frac{10}{27}\frac{\alpha^4}{n^8}-\frac{17}{27}\frac{\alpha^6}{n^{12}}+\cdots \right]\,,
	\end{align} 
	and the corresponding normalized generating function defined in \eqref{eq: generating function} reads
	\begin{align}\label{fn gravity result}
		f_{n}(\alpha):=\frac{\mathcal{Z}_n(\alpha)}{\mathcal{Z}_n(0)}=e^{\frac{cl}{6n}\left(-\frac{1}{3}\frac{\alpha^2}{n^4}+\frac{10}{27}\frac{\alpha^4}{n^8}-\frac{17}{27}\frac{\alpha^6}{n^{12}}+\cdots \right)} \,.
	\end{align}
	\eqref{fn gravity result} is the final result from higher spin holography which will be reproduced by our CFT calculations in \secref{sec:4}.
	
	

	\section{CFT calculation}\label{sec:4}
	Even though it is customary to employ charged twist fields in the computation of $U(1)$ charged moments as discussed in \cite{GoldsteinSela}, it is not clear whether this method still works in the higher spin case, due to the non-linearity of the $W_3$ algebra. Therefore, in order to perform the CFT calculation for the charged moments, an alternative perspective to that of the charged twist fields is needed.
	
	In this section, we focus on the $\alpha$-dependent part of the charged moments \eqref{eq: charged moments}, which is given by the normalized generating function $f_{n}(\alpha)$ defined in \eqref{eq: generating function}. Following \cite{Gaberdiel:2012yb,Datta:2014ska,Datta:2014uxa,Datta:2014zpa}, we present a perturbative calculation for $f_{n}(\alpha)$, which in principle can be computed up to arbitrary orders of $\alpha$. Our basic idea is to directly treat the operator $e^{2\pi i\alpha \mathcal{Q}_{\mathcal{A}}}$ as a defect line operator in the replica manifold. We first show that the perturbative expansion of the charged moments can be expressed as the connected correlators of the modular charge $\mathcal{Q}_{\mathcal{A}}$. 
	We compute the result up to $\alpha^4$ order, which matches with the gravity result \eqref{fn gravity result}. As another example of our perturbative method, in \appref{appendix: C}, we show that the usual R\'enyi entropy can also be obtained from this defect line perspective, without using the replica trick.
	
	Our perturbative method provides an efficient way to compute the charged moments. We expect that our method can be applied to CFTs with other types of extended conformal symmetries. In addition, since our calculation is only based on the symmetry algebra, this indicates that the universal behavior of the charged moments is determined by the symmetry.
	
	\subsection{Perturbative expansion of charged moments}\label{sec:4.1}
	
	We consider the normalized generating function $f_{n}(\alpha)$ for a single entangling interval. As in \secref{sec: 3.1}, we neglect the contribution of the boundary entropy in the BCFT perspective, and only consider the vacuum state contribution to the charged moments. Since the boundary state in the original complex $z$-plane is represented by some non-trivial operator content at the endpoints of the interval $\mathcal{A}$, neglecting the boundary entropy indicates that we only consider the contribution from the identity operator\footnote{We assume that the boundary condition is elementary, i.e. that it contains the representation of the identity field exactly once, therefore the vacuum is non-degenerate.}. Under this assumption, the result for $f_{n}(\alpha)$ is nothing but the vacuum expectation value of the defect line operator $e^{2\pi i \alpha \mathcal{Q}_{\mathcal{A}}}$ on the replica $n$-fold,
	\begin{align}
		f_{n}(\alpha)=\frac{\mathcal{Z}_{n}(\alpha)}{\mathcal{Z}_{n}(0)}=\frac{\Tr_{\mathcal{H}_{ab}}\left[\rho_{\mathcal{A}}^n e^{2\pi i \alpha \mathcal{Q}_{\mathcal{A}}}\right]}{\Tr_{\mathcal{H}_{ab}}\left[\rho_{\mathcal{A}}^n\right]}\approx \left\langle e^{2\pi i \alpha \mathcal{Q}_{\mathcal{A}}}\right\rangle_{n}\ ,
	\end{align}
	where $\langle\cdots\rangle_{n}$ denotes the vacuum expectation value on the replica $n$-fold.

	Since, in this case, the topology of the corresponding replica $n$-fold is still a Riemann sphere, one can map the system to a flat complex plane with coordinate $u$ by the standard uniformization mapping
	\begin{align}\label{uni mapping}
		u=\left(\frac{z-z_2}{z_1-z}\right)^\frac{1}{n}\ .
	\end{align}
	Under the conformal mapping \eqref{uni mapping}, the spin 3 modular charge \eqref{eq: modular Q} transforms as
	\begin{align}\label{eq: modular Q replica plane}
		\mathcal{Q}_{\mathcal{A}}=-\int_{\delta}^{\frac{1}{\delta}}\frac{du}{2\pi i}\xi^u\xi^u W_{uuu}+\cc=\int_{\delta}^{\frac{1}{\delta}}\frac{du}{2\pi i}\frac{u^2}{n^2}W(u)+\cc\ ,
	\end{align}
	with
	\begin{align}
		\xi^u=\xi^z \left(\frac{du}{dz}\right)=-\frac{iu}{n}\ ,
	\end{align}
	The cut-offs $\delta$ and $1/\delta$ around $u=0$ and $u=\infty$ are associated with the cut-offs around $z=z_2$ and $z=z_1$ by the mapping \eqref{uni mapping}, and given by
	\begin{align}\label{eq: cutoff delta}
		\delta=\left(\frac{\epsilon}{z_1-z_2}\right)^{\frac{1}{n}}\,.
	\end{align}
	A perturbative expansion of the generating function around $\alpha=0$ yields
	\begin{align}\label{fn expansion}
		\log f_n(\alpha)
		&=\log f_n(0)+\alpha\frac{f_n^{(1)}(0)}{f_n(0)}+\frac{\alpha^2}{2!}\left(\frac{f_n^{(2)}(0)f_n(0)-(f^{(1)}_n(0))^2}{f_n(0)^2}\right)+\cdots\nonumber\\
		&=\sum_{k=1}^{\infty}\frac{\alpha^k}{k!}\left\langle (2\pi i \mathcal{Q}_{\mathcal{A}})^k\right\rangle_{n, c}\,,
	\end{align}
	where $\langle\cdots\rangle_{n, c}$ denotes the connected correlator on the replica $n$-fold, which is the complex $u$- plane. In the last step of \eqref{fn expansion}, we used
	\begin{align}
		\frac{f^{(k)}_n(0)}{f_n(0)}=\left\langle (2\pi i \mathcal{Q}_{\mathcal{A}})^k\right\rangle_{n}\,,
	\end{align}
	as well as
	\begin{align}
		\frac{f^{(1)}_n(0)}{f_n(0)}=\left\langle (2\pi i \mathcal{Q}_{\mathcal{A}})\right\rangle_{n }=0\ ,
	\end{align}
	which holds due to the vanishing one point function in the complex  $u$ plane, $\langle W(u) \rangle=0$. In order to solve \eqref{fn expansion}, the connected correlator of spin 3 currents needs to be derived. In this process, the $WW$ OPE,
	\begin{align}\label{ww ope}
		W(u)W(0)&\sim \frac{5c}{6}\frac{1}{u^6}+\frac{5T(0)}{u^4}+\frac{5}{2}\frac{\partial T(0)}{u^3}\nonumber\\
		&+\frac{1}{u^2}\left(\frac{16}{c}\Lambda(0)+\frac{3}{4}\partial^2T(0)\right)+\frac{1}{u}\left(\frac{8}{c}\partial\Lambda(0)+\frac{1}{6}\partial^3T(0)\right)\,,
	\end{align}
	with the nonlinear term $\Lambda(u)$ given by
	\begin{align}\label{lambda}
		\Lambda(u)=:T(u)T(u):-\frac{3}{10}\partial^2 T(u)\,,
	\end{align}
	is of crucial importance. The contribution from the nonlinear term $\Lambda$ in \eqref{ww ope} for correlation functions of spin 3 currents is suppressed in the large $c$ limit\footnote{The nonlinear term will contribute to the leading order of the correlation functions, if heavy operators, whose conformal weight and spin 3 charge scales in $c$, are present in the background. Since we are considering the vacuum background, this is not the case here}. After neglecting the nonlinear terms in \eqref{ww ope}, a recursion relation for the connected correlation functions of spin 3 currents is found,
	\begin{align}\label{Recursion}
		&\langle{W(v)W(u_1)\cdots W(u_{k-1})}\rangle_c\nonumber\\ =&\sum_{i=1}^{k-1}F(v,u_i;\partial_{u_i})\langle W(u_1)\cdots T(u_i)\cdots W(u_{n-1})\rangle_c\nonumber\\
		=&\sum_{i=1}^{k-1}\sum_{j\neq i}F(v,u_i;\partial_{u_i})G(u_i,u_j;\partial_{u_j})\langle W(u_1)\cdots W(u_{i-1})W(u_{i+1})\cdots W(u_{n-1})\rangle_c\,,
	\end{align}
	with
	\begin{align}
		F(v,u_i;\partial_{u_i})&=\frac{5}{(v-u_i)^4}+\frac{5\partial_{u_i}}{2(v-u_i)^3}+\frac{3\partial_{u_i}^2}{4(v-u_i)^2}+\frac{\partial^3_{u_i}}{6(v-u_i)}\ ,\nonumber\\
		G(u_i,u_j;\partial_{u_j})&=\frac{3}{(u_i-u_j)^2}+\frac{\partial_{u_j}}{u_i-u_j}\ .
	\end{align}
	The result \eqref{Recursion} is a recursion relation between a $k$-point function and a $k-2$-point function, with $k\ge 3$. Following \eqref{Recursion}, two points are concluded: Firstly, since the two point function $\langle W(v)W(u_1) \rangle$ scales as $c$, all the even higher point connected correlators also scale as $c$. Furthermore, the vanishing of the one point function implies that all the odd point connected correlators vanish. Hence, the expansion of the generating function $f_n(\alpha)$ in \eqref{fn expansion} simplifies to
	\begin{align}\label{fn expansion 2}
		\log f_n(\alpha)=\sum_{k=1}^{\infty}\frac{\alpha^{2k}}{(2k)!}\left\langle (2\pi i \mathcal{Q}_{\mathcal{A}})^{2k}\right\rangle_{n, c}\,.
	\end{align}
	As a consequence of focussing on the leading $\log \ell/\epsilon$ piece in the charged moments, the choice for the automorphism $\Omega$ in $\mathcal{Q}_{\mathcal{A}}$ does not affect the result for $f_{n}(\alpha)$.
	Moreover, observing that $\mathcal{Q}_{\mathcal{A}}$ in \eqref{eq: modular Q replica plane} in the $u$-coordinates naturally includes a $1/n^2$ factor, it is evident that the form of \eqref{fn expansion 2} already matches that of the gravity result \eqref{fn gravity result}.
	
	\subsection{CFT calculation for charged moments}\label{sec:4.2}
	In this section, we calculate the CFT charged moments by computing the connected correlator of modular charges in \eqref{fn expansion 2} up to forth order in the chemical potential, and verify that the result matches the gravity result \eqref{fn gravity result}.
	
	We start from the quadratic order correlator of the modular charges. Employing the definition \eqref{eq: modular Q replica plane}, it is straightforward to compute 
	\begin{align}\label{2 order}
		\left\langle(2\pi i \mathcal{Q}_{\mathcal{A}} )^{2}\right\rangle_{n, c}
		&=\frac{1}{n^4}\int_{\delta}^{\frac{1}{\delta}}du_1\int_{\delta}^{\frac{1}{\delta}}du_2\ u_1^2\ u_2^2 \ \langle W(u_1)W(w_2)\rangle_{n,c}+\cc\nonumber\\
		&=\frac{1}{n^4}\int_{\delta}^{\frac{1}{\delta}}du_1\int_{\delta}^{\frac{1}{\delta}}\ \frac{5c}{6}\frac{u_1^2\ u_2^2}{(u_1-u_2)^6}+\cc\nonumber\\
		&\approx-\frac{1}{n^4}\int_{\delta}^{1/\delta}du_1 \ \frac{c}{36u_1}+\cc\nonumber\\
		&=-\frac{cl}{9n^5}\,,
	\end{align}
	where we used the relation \eqref{eq: cutoff delta} in the last step. Moving on to the forth order, one can use the recursion relation \eqref{Recursion} to obtain the connected four point  correlation function.\footnote{The expression for the connected four point correlator is very long, but can be found in the \texttrademark\textit{Mathematica} notebook supplementing the ArXiv version of this paper.} By an analogous calculation as for the quadratic order, one finds
	\begin{align}\label{4 order}
		\left\langle(2\pi i \mathcal{Q}_{\mathcal{A}})^4\right\rangle_{n, c}
		&=\frac{1}{n^8}\left(\prod_{i=1}^4\int_{\delta}^{\frac{1}{\delta}}du_{i}\ u_i^2\right)\langle W(u_1)W(u_2)W(u_3)W(u_4)\rangle_{c}
		=\frac{40cl}{27n^9}\ .
	\end{align}
	Higher order are worked out analogously via the recursion relation. 	Inserting these results into \eqref{fn expansion 2}, the generating function up to $\alpha^4$ is obtained,
	\begin{align}
		\log f_{n}(\alpha)=\frac{cl}{6n}\left(-\frac{1}{3}\frac{\alpha^2}{n^4}+\frac{10}{27}\frac{\alpha^4}{n^8}+\cdots \right) \ .
	\end{align}
	As advertized this matches exactly the gravity result \eqref{fn gravity result} derived from the partition function of the topological spin 3 black hole. Higher orders can in principle be calculated in the same manner, only with the corresponding integrals becoming somewhat harder to perform.	
	
	
\section{Conclusion}\label{sec:conclusion}
	
	In this work, we employed the $W_3$ algebra acting on the entanglement interval in the CFT to give a new and independent CFT derivation of the charged moments in the AdS$_3$/CFT$_2$ correspondence based on $SL(3,\mathbb{R})\times SL(3,\mathbb{R})$ higher spin gravity. To derive this entanglement $W_3$ algebra, we defined a modular charge operator on the entangling interval in the same way as the modular Hamiltonian is defined: It is the dual to the corresponding chemical potential in the charged moment, which is nothing but the grand canonical partition function associated with the entangling interval. Based on this interpretation as a grand canonical partition function, we moreover identify the charged moments with the partition function of a charged topological black hole \cite{MyersCasiniBlanco}. 
	
	In our calculation, the entanglement $W_3$ algebra was used to 
	evaluate the charged moments by reducing it order by order to connected  correlation functions of modular charge operator insertions. 
	 While the reduction of the charged moments to modular charge correlators is possible in principle to an arbitrarily high order in an expansion in the chemical potential, the actual reduction using the $W_3$ algebra became technically more and more involved at higher orders in this expansion due to the non-linear of the $W_3$ algebra. We calculated up to quartic order in the expansion in the spin 3 chemical potential, and show that the corresponding charged moment matches with known results from the higher spin gravity side of the AdS$_3$/CFT$_2$ correspondence. Our CFT calculation hence constitutes a new independent check of the AdS$_3$/CFT$_2$ correspondence, which in principle also can be extended to higher orders using e.g. computer algebra methods.

	One important insight developed in this work is that equipartition of entanglement fails to hold at leading order in the large $c$ expansion in the $W_3$ CFT dual to spin 3 gravity. For free fields \cite{Murciano:2020vgh,Murciano:2020lqq, Bonsignori:2020laa, Capizzi:2020jed, tan2020particle, calabrese2021symmetry}, equipartition of entanglement was found to be present for all the terms in the symmetry resolved entanglement entropy, and a breakdown of equipartition was observed for WZW models to subleading order in the large interval length expansion in \cite{calabrese2021symmetry}. In our case, the breakdown of equipartition to leading order in large $c$ is a consequence of the structure of the $W_3$ algebra, which allows higher connection correlation functions $\langle W^n\rangle_c$ of the spin 3 modular charge $W$ to be non-vanishing. On the contrary, free bosons for example are governed by the $U(1)$ Kac-Moody algebra, in which case the theory is Gaussian, and the higher connected correlators of the conserved current always vanish. The same holds for the WZW case of \cite{calabrese2021symmetry}, where the algebra is non-abelian. The breakdown of equipartition at leading order in large $c$ is interesting in its own right, but in particular also hints to an interesting charge substructure in how entanglement is encoded in the dual space-time geometry for higher spin holography. It would be interesting to further investigate this in the context of e.g. the bulk reconstruction program \cite{VanRaamsdonk,Hamilton:2005ju,Melnikov:2018fhb,Harlow:2018jwu,Harlow:2018tng,Abt:2017pmf,Abt:2018ywl}.	
	
Another interesting question about the higher spin charged moments arises from its nontrivial phase structure from the gravity perspective. As mentioned in \secref{sec: 3.2}, the higher spin black hole solutions obtained via the holonomy condition contain different branches. The dominance of different branches is controlled by the dimensionless parameter $\alpha/n^2$. Since our perturbative calculation of the charged moments in CFT breaks down for large enough $\alpha$, we only focused on the BTZ branch in the regime $\alpha/n^2\ll 1$. For the non-perturbative analysis of the higher spin charged moments, it is necessary to understand how the holonomy condition arises from the CFT perspective. As shown in \cite{deBoer:2014sna}, if only local primary operators are present in the correlation function, the holonomy condition for the $SL(3,\mathbbm{R})$ Chern-Simons connection can be identified with the trivial monodromy condition imposed on the corresponding conformal partial wave function in the dual $W_3$ CFT. This mechanism is based on the null state equation for the correlation functions of $W_3$ primaries, as well as the assumption of the dominance of the $W_3$ vacuum conformal block, c.f.~\cite{fitzpatrick2014universality,Gerbershagen:2021yma} for a detailed discussion on the  monodromy method in Virasoro case, as well as the generalization for $W_N$ symmetry \cite{Karlsson:2021mgg}.  Since in the $W_3$ case, the nonlocal defect operator $e^{2\pi i \alpha \mathcal{Q}_{\mathcal{A}}}$ does not readily factorize into local vertex operators, a generalization of the null state equation to the case of nonlocal operator insertions seems necessary and may provide an approach for solving the higher spin charged moments in the non-perturbative regime $\alpha/n^2\to \infty$.
	
	Furthermore, it is also interesting to investigate the higher spin charged moments from the perspective of Toda theory. The usual R\'enyi entropy defined on the $n$-sheeted branched covering of a single interval is captured by the Liouville action \cite{Hung:2011nu},
	\begin{align}\label{Liouville}
		\langle \sigma_{n}(z_1)\tilde{\sigma}_n(z_2)\rangle=e^{-S_{Liouville}[\phi]}\,,
	\end{align}
	where the Liouville field $\phi$ is induced by the conformal mapping $f(z)$ from the $n$-sheeted covering space to a complex plane, given by $e^{-2\phi}=|\partial f(z)|^2$. For the higher spin charged moments, in addition to the conformal transformation $f(z)$ which contributes to the Liouville action, one should also include effect of the finite spin 3 transformation induced by the operator $e^{2\pi i\alpha \mathcal{Q}_{\mathcal{A}}}$. The finite spin $3$ transformation has been discussed in two-dimensional BF theory, and was used to construct the effective action for the higher spin SYK model \cite{Gonzalez:2018enk}. The question is then what is the higher spin analog of Liouville action in \eqref{Liouville} in the presence of the defect line $e^{2\pi i\alpha \mathcal{Q}_{\mathcal{A}}}$. We suspect the resulting effective action to be that of Toda field theory.
	
	Finally, since the decomposition of the entanglement entropy into symmetry-resolved pieces is universal, it would be interesting to understand the symmetry resolution of bulk entanglement in AdS/CFT, as well as the physical interpretation thereof. In \cite{Lewkowycz:2013nqa} it was shown that entanglement of bulk fields across the Ryu-Takayanagi surface yields the universal subleading correction to the large $c$ limit of holographic entanglement entropy. For Einstein-Maxwell-Chern-Simon theory, also considered in \cite{zhao2021symmetry,weisenberger2021symmetry}, bulk entanglement was argued in  \cite{belin2013holographic} to be responsible for the $c \rightarrow c+1$ shift of the central charge due to the Sugawara energy momentum tensor. In that case, the bulk entanglement was traced back to the contributions of boundary photons along the RT surface, in complete analogy to the boundary photons along the UV AdS cutoff surface. Einstein-Hilbert-Chern-Simons theory may hence serve a fruitful starting point to understand bulk entanglement symmetry resolution, after which one could understand $SL(N,\mathbbm{R})$ in a second step.  Relatedly, it will also be interesting to investigate the symmetry resolution of the island contribution to the entanglement entropy \cite{Engelhardt:2014gca}, in particular whether different charge sectors contribute distinctively different to black hole evaporation or not. Since both bulk entanglement and the quantum contributions to the Page curve are quantum gravity effects, understanding their symmetry resolution might provide new perspective on the structure of quantum gravity itself.
	
	
	\acknowledgments
	We thank Shira Chapman, Daniel Grumiller, Stefan Fredenhagen, Giuseppe Di Giulio and Sara Murciano for useful discussions. R.M., C.N. and S.Z. acknowledge support by the Deutsche Forschungsgemeinschaft (DFG, German Research Foundation) under Germany's Excellence Strategy through the W\"urzburg‐Dresden Cluster of Excellence on Complexity and Topology in Quantum Matter ‐ ct.qmat (EXC 2147, project‐id 390858490). The work of R.M. and C.N. was furthermore supported via project id 258499086 - SFB 1170 ’ToCoTronics’. S.Z. is financially supported  by the China Scholarship Council.

	\begin{appendix}

		\section{Conventions}\label{appendix A}
		In the main text, the $W_3$ algebra followed from the variation law \eqref{eq: transformation law 1}. In modes this algebra becomes
		\begin{align}
			&\left[ L_m, L_n\right]=(m-n)L_{m+n}+\frac{c}{12}m(m^2-1)\delta_{m+n,0}\ ,\nonumber\\
			&\left[L_m, W_n\right]=(2m-n)W_{m+n}\ ,\nonumber\\
			&\left[W_m, W_n\right]=-\frac{1}{12}(m-n)(2m^2+2n^2-mn-8)L_{m+n}\nonumber\\
			&\quad \quad\quad \quad \quad\quad
			+\frac{8}{c}(m-n)\lambda_{m+n}+\frac{5c}{6}\frac{1}{5!}m(m^2-1)(m^2-4)\delta_{m+n,0}\ ,
		\end{align}
		with
		\begin{align}
			\lambda_m=\sum_{n}L_{n}L_{m-n}-\frac{3}{10}(m+3)(m+2)L_m\ .
		\end{align}
		The mode expansions of the currents are given by
		\begin{align}
			T(z)=\sum_{n}\frac{L_n}{z^{n+2}}\ ,\quad W(z)=\sum_{n}\frac{W_n}{z^{n+3}}\ ,
		\end{align}
		and the OPEs read
		\begin{align}
			&T(z)T(0)\sim\frac{c/2}{z^2}+\frac{2T(0)}{z^2}+\frac{\partial T(0)}{z}\ ,\\
			&T(z)W(0)\sim\frac{3W(0)}{z^2}+\frac{\partial W(0)}{z}\ ,\\
			&W(z)W(0)\sim \frac{5c}{6}\frac{1}{z^6}+\frac{5T(0)}{z^4}+\frac{5}{2}\frac{\partial T(0)}{z^3}\nonumber\\
			&\quad \quad\quad\quad\quad\quad+\frac{1}{z^2}\left(\frac{16}{c}\Lambda(0)+\frac{3}{4}\partial^2T(0)\right)+\frac{1}{z}\left(\frac{8}{c}\partial\Lambda(0)+\frac{1}{6}\partial^3T(0)\right)\,,
		\end{align}
		where the nonlinear term $\Lambda(z)$ is defined as
		\begin{align}
			\Lambda(z)=:T(z)T(z):-\frac{3}{10}\partial^2 T(z) \ .
		\end{align}
		\section{Entanglement $W_3$ algebra}\label{app: W3algebra}
		In this section, we construct a single copy $W_3$ algebra on the single entangling interval $\mathcal{A}$ by imposing symmetry-preserving boundary conditions at the cut-off circles around the endpoints of the interval $\mathcal{A}$. The construction for the modes of the algebra is based on the BCFT description of entanglement \cite{OhmoriTachikawa, CardyTonni, hu2020emergent}, in which, one transforms the original $z$-complex plane to a semi-annulus embedded in the upper-half $u$-complex plane, and the boundary conditions around the original cut-off circles become the standard conformal boundary conditions on the real axis of the upper-half $u$-complex plane. By a straightforward calculation, we show that the commutator of the $W_3$ modes defined on the interval $\mathcal{A}$ indeed furnishes a single copy of the $W_3$ algebra, with the zero modes of the algebra proportional to the modular Hamiltonian and spin 3 modular charge up to some constant factors.
		
		We consider a single interval $A$ on the $z$-complex plane and locate its endpoints, without loss of generality, at $z=z_1$ and $z=z_2$, with $|z_1|\ge|z_2|$. In order to regularize the endpoints, we impose cut-off circles around each endpoint such that the region we considered here satisfies $|z-z_1|\ge |\epsilon|$ and $|z-z_2|\ge |\epsilon|$. This region can be mapped to a semi-annulus on a upper-half $u$-complex plane by the following conformal transformation
		\begin{align}\label{transformation u-z}
			u=e^{i\theta(z)}\ ,
		\end{align}
		with
		\begin{align}
			\theta(z)=\frac{\pi}{2}-\frac{\pi}{2l}\log\left(\frac{z-z_2}{z_1-z}\right)\ , \quad l=\log\left(\frac{z_1-z_2}{\epsilon}\right)\in \mathbb{R}\ .
		\end{align}
		The inner and outer semi-circles are identified\footnote{This identification turns the semi-annulus in the complex $u$-plane into an honest annulus. To avoid confusion, we keep referring to it as the semi-annulus however.}, and they correspond to the interval $\mathcal{A}$ in the original $z$-complex plane. The cut-off circles around the endpoints in the $z$-complex plane are now mapped to the two boundaries of the semi-annulus located along the real axis of the $u$-complex plane.
		
		Since the semi-annulus is embedded in the upper-half $u$-plane, one can implement the BCFT description by imposing conformal boundary conditions at the real axis of the $u$-complex plane, which is given by
		\begin{align}\label{eq: bc T}
			T(u)=\bar{T}(\bar{u})\  ,\quad  \forall\ u=\bar{u}\in \mathbb{R}\ .
		\end{align}
		The physical meaning of the boundary condition \eqref{eq: bc T} is that the energy flow is reflected at the real axis of the $u$-complex plane, and as a result, the original two copy of Virasoro modes are not independent any more when acting on the boundary states.
		
		Due to the presence of the spin 3 currents in the theory, we can further specify the following maximally symmetric boundary condition for spin 3 current such that it is compatible with \eqref{eq: bc T},
		\begin{align}\label{eq: bc W}
			W(u)=\Omega \bar{W}(\bar{u})\ , \quad \forall \ u=\bar{u}\in \mathbb{R}\ .
		\end{align}
		where $\Omega$ denotes the automorphism for the spin 3 current, given by $\Omega=\pm\id$.
		
		The conformal boundary conditions \eqref{eq: bc T} and \eqref{eq: bc W} allow to identify $\bar{T}(\bar{u})$ and $\Omega\bar{W}(\bar{u})$ as the analytic continuation of $T(u)$ and $W(u)$ in the lower-half $u$-complex plane. Hence, it is easy to see that the Virasoro and spin 3 modes defined as
		\begin{align}\label{eq: u-plane modes}
			L^{(\mathcal{A})}_n&=\frac{1}{2\pi i}\int_{C} du \ u^{n+1} T(u)-\frac{1}{2\pi i}\int_{C}d\bar{u}\ \bar{u}^{n+1}\bar{T}(\bar{u})\ ,\nonumber\\
			W^{(\mathcal{A})}_n&=\frac{1}{2\pi i}\int_{C} du \ u^{n+2} W(u)-\frac{1}{2\pi i}\int_{C}d\bar{u}\ \bar{u}^{n+2}\ \Omega\bar{W}(\bar{u})\ .
		\end{align}
		furnish a single copy of $W_3$ algebra for the BCFT. Here the integration contour $C$ is a semi-circle going counterclockwise around the origin.
		
		By the conformal transformation \eqref{transformation u-z}, one can rewrite \eqref{eq: u-plane modes} as
		\begin{align}
			L^{(\mathcal{A})}_n&=-\frac{1}{2\pi i}\int_{z_2+\epsilon}^{z_1-\epsilon} dz \left(\frac{du}{dz}\right) u(z)^{n+1} \left(\frac{du}{dz}\right)^{-2}\left(T(z)-\frac{c}{12}\{u,z\}\right)+\cc\nonumber\\
			&=\frac{c}{24}\left(1+\frac{4l^2}{\pi^2}\right)\delta_{n,0}+\frac{1}{2\pi}\int_{z_2+\epsilon}^{z_1-\epsilon} dz\ \frac{e^{i n \theta(z)}T(z)}{\partial_z\theta(z)}+\frac{1}{2\pi}\int_{\bar{z}_2+\bar{\epsilon}}^{\bar{z}_1-\bar{\epsilon}}d\bar{z}\ \frac{e^{-in\bar{\theta}(\bar{z})}\bar{T}(\bar{z})}{\partial_{\bar{z}}\bar{\theta}(\bar{z})}\ ,\label{eq: modes L z-plane}\\
			W^{(\mathcal{A})}_n&=-\frac{1}{2\pi i}\int_{z_1+\epsilon}^{z_1-\epsilon} dz \left(\frac{du}{dz}\right) u(z)^{n+2} \left(\frac{du}{dz}\right)^{-3} W(z) +\cc\nonumber\\
			&=\frac{1}{2\pi i}\int_{z_2+\epsilon}^{z_1-\epsilon}dz \ \frac{e^{i n\theta(z)}W(z)}{\left(\partial_z \theta(z)\right)^2}-\frac{1}{2\pi i}\int_{\bar{z}_2+\bar{\epsilon}}^{\bar{z}_1-\bar{\epsilon}} d\bar{z} \frac{e^{-i n \bar{\theta}(\bar{z})}\ \Omega\bar{W}(\bar{z})}{\left(\partial_{\bar{z}}\bar{\theta}(\bar{z})\right)^2}\ ,\label{eq: modes W z-plane}
		\end{align}
		where the minus signs in the first lines of \eqref{eq: modes L z-plane} and \eqref{eq: modes W z-plane} appear since we change the direction of the integration contour $C$.
		The boundary conditions on the $z$-complex plane are now defined at the two cut-off circles, i.e. $|z-z_1|=|\epsilon|$ and $|z-z_2|=|\epsilon|$, given by
		\begin{align}\label{bc z plane 1}
			0&=\left(\frac{du}{dz}\right)^{-2}\left(T(z)-\frac{c}{12}\{u,z\}\right)-\left(\frac{d\bar{u}}{d\bar{z}}\right)^{-2}\left(\bar{T}(\bar{z})-\frac{c}{12}\{\bar{u},\bar{z}\}\right)\nonumber\\
			&=-\frac{T(z)}{u(z)^2 \left(\partial_z \theta(z)\right)^2}+\frac{\bar{T}(\bar{z})}{\bar{u}(\bar{z})^2\left(\partial_{\bar{z}}\bar{\theta}(\bar{z})\right)^2}+\frac{c}{24}\left(1+\frac{4l^2}{\pi^2}\right)\left(\frac{1}{u(z)^2}-\frac{1}{\bar{u}(\bar{z})^2}\right)\ ,
		\end{align}
		and
		\begin{align}\label{bc z plane 2}
			0=\left(\frac{du}{dz}\right)^{-3}W(z)-\left(\frac{d\bar{u}}{d\bar{z}}\right)^{-3} \Omega \bar{W}(\bar{z})=\frac{i W(z)}{u(z)^3\left(\partial_z \theta(z)\right)^3}+\frac{i \ \Omega \bar{W}(\bar{z})}{\bar{u}(\bar{z})^3\left(\partial_{\bar{z}}\bar{\theta}(\bar{z})\right)^3}\ .
		\end{align}
		Notice that at the cut-off circles, we have $u(z)=\bar{u}(\bar{z})\in \mathbb{R}$. Therefore, the equations \eqref{bc z plane 1} and \eqref{bc z plane 2} further reduce to
		\begin{align}\label{bc z plane final}
			\frac{T(z)}{\left(\partial_z \theta(z)\right)^2}=\frac{\bar{T}(\bar{z})}{\left(\partial_{\bar{z}}\bar{\theta}(\bar{z})\right)^2}\ ,\quad \frac{W(z)}{\left(\partial_z \theta(z)\right)^3}=-\frac{\Omega\bar{W}(\bar{z})}{\left(\partial_{\bar{z}}\bar{\theta}(\bar{z})\right)^3}\ .
		\end{align}
		with $|z-z_1|=|\epsilon|$ or $|z-z_2|=|\epsilon|$.
		
		Using \eqref{eq: modes L z-plane} and \eqref{eq: modes W z-plane} as well as the boundary conditions \eqref{bc z plane final}, it is possible to verify that the modes defined in \eqref{eq: modes L z-plane} and \eqref{eq: modes W z-plane} indeed form a $W_3$ algebra by a straightforward calculation. Here we present the example for calculating $\left[L^{(\mathcal{A})}_n, L^{(\mathcal{A})}_m\right]$, which reads
		\begin{align}\label{algebra calculation 1}
			[L^{(\mathcal{A})}_n, L^{(\mathcal{A})}_m]&=\int_{z_2+\epsilon}^{z_1-\epsilon}\frac{dz'}{2\pi}\int_{z_2+\epsilon}^{z_1-\epsilon}\frac{dz}{2\pi}\frac{e^{i n \theta(z')}e^{i m \theta(z)}T(z')T(z)}{\partial_{z'}\theta(z')\partial_{z}\theta_(z)}\nonumber\\
			&\quad 
			-\int_{z_2+\epsilon}^{z_1-\epsilon}\frac{dz}{2\pi}\int_{z_2+\epsilon}^{z_1-\epsilon}\frac{dz'}{2\pi}\frac{e^{i n \theta(z')}e^{i m \theta(z)}T(z)T(z')}{\partial_{z'}\theta(z')\partial_{z}\theta_(z)}+\cc\nonumber\\
			&=\int_{z_2+\epsilon}^{z_1-\epsilon}\frac{dz}{2\pi}\ointclockwise_{z'=z}\frac{dz'}{2\pi}\frac{e^{i n \theta(z')}e^{i m \theta(z)}}{\partial_{z'}\theta(z')\partial_{z}\theta_(z)}\left(\frac{c/2}{(z'-z)^4}+\frac{2T(z)}{(z'-z)^2}+\frac{\partial_z T(z)}{z'-z}\right)+\cc\nonumber\\
			&=\int_{z_2+\epsilon}^{z_1-\epsilon}dz \ \left\{ -\frac{ c}{24\pi}n\left(n^2+\frac{4l^2 }{\pi^2}\right)\partial_z\theta(z)e^{i(n+m)\theta(z)}+\frac{(n-m)e^{i(n+m)\theta(z)}T(z)}{2\pi \partial_z \theta(z)}\right\} \nonumber\\
			& \quad + \int_{\bar{z}_2+\bar{\epsilon}}^{\bar{z}_1-\bar{\epsilon}}d\bar{z}\ \left\{-\frac{c}{24\pi}n\left(n^2+\frac{4l^2}{\pi^2}\right)\partial_{\bar{z}}\bar{\theta}(\bar{z})e^{-i(n+m)\bar{\theta}(\bar{z})}+\frac{(n-m)e^{-i(n+m)\bar{\theta}(\bar{z})}\bar{T}(\bar{z})}{2\pi \partial_{\bar{z}} \bar{\theta}(\bar{z})}\right\}\nonumber\\
			&\quad -\frac{i u(z)^{m+n}T(z)}{2\pi\left(\partial_z\theta(z)\right)^2}\Bigg|_{z_2+\epsilon}^{z_1-\epsilon}+\frac{i\bar{u}(\bar{z})^{m+n}\bar{T}(\bar{z})}{2\pi\left(\partial_{\bar{z}}\bar{\theta}(\bar{z})\right)^2}\Bigg|_{\bar{z}_2+\bar{\epsilon}}^{\bar{z}_1-\bar{\epsilon}}\nonumber\\
			&=\frac{c}{12}n(n^2+\frac{4l^2}{\pi^2})\delta_{n+m,0}+\frac{(n-m)}{2\pi}\left\{\int_{z_2+\epsilon}^{z_1-\epsilon} dz\ \frac{e^{i (n+m) \theta(z)}T(z)}{\partial_z\theta(z)}+\int_{\bar{z}_2+\bar{\epsilon}}^{\bar{z}_1-\bar{\epsilon}}d\bar{z}\ \frac{e^{-i(n+m)\bar{\theta}(\bar{z})}\bar{T}(\bar{z})}{\partial_{\bar{z}}\bar{\theta}(\bar{z})}\right\}\nonumber\\
			&=(n-m)L^{(\mathcal{A})}_{n+m}+\frac{c}{12}n(n^2-1)\delta_{n+m,0}\,,
		\end{align}
		where in the forth step of \eqref{algebra calculation 1}, we used the boundary condition \eqref{bc z plane final} as well as $u(z)=\bar{u}(\bar{z})\in\mathbb{R}$ at the two cut-off circles. Other commutators of the entanglement $W_3$ algebra can be checked in a similar way.
		
		\section{R\'enyi entropy via defect line perspective}\label{appendix: C}
		In this appendix, we provide a perturbative calculation for the ordinary R\'enyi entropy from the perspective of the defect line operator. We start by considering the following quantity
		\begin{align}\label{eq: gn}
			g(n)=\Tr_{\mathcal{H}_{ab}}[\rho_{\mathcal{A}}^{1+n}]\approx\langle e^{-2\pi n \mathcal{H}_{\mathcal{A}}}\rangle=e^{-2\pi n C}\langle e^{-2\pi n H_0}\rangle\,,
		\end{align}
		where we again neglect the boundary entropy contribution, and $H_0$ is given by
		\begin{align}
			H_0=\mathcal{H}_{\mathcal{A}}-C=-\int_{z_2+\epsilon}^{z_1-\epsilon}\frac{dz}{2\pi i}\xi^z T_{zz}+\int_{\bar{z_2}+\bar{\epsilon}}^{\bar{z_1}-\bar{\epsilon}}\frac{d\bar{z}}{2\pi i}\xi^{\bar{z}}T_{\bar{z}\bar{z}}\,.
		\end{align}
		The normalization constant C defined in \eqref{eq: modular H} is given by $C=\frac{cl}{6\pi}$, which is fixed by the thermal partition function $Z_{1}(0)=e^{\frac{cl}{6}}$ as well as the definitions for $L^{(\mathcal{A})}_{0}$ and $\mathcal{H}_{\mathcal{A}}$ in \eqref{eq: zero mode L0} and \eqref{eq: modular H}.
		
		Note that the equation \eqref{eq: gn} is nothing but the vacuum expectation value of the defect operator $e^{-2\pi n \mathcal{H}_{\mathcal{A}}}$ in the original $z$-complex plane, where no replica construction is employed. Analogous to the analysis in \secref{sec:4.1}, the perturbative expansion of $\log g(n)$ with respect to $n$ is given by
		\begin{align}\label{eq: log gn}
			\log g(n)=-\frac{ncl}{3}+\sum_{k=2}^{\infty}\frac{n^k}{k!}\langle(-2\pi H_0)^k\rangle_c\,,
		\end{align}
		where $\langle \cdots\rangle_{c}$ denotes the connected correlator in the $v$-complex plane. By the conformal mapping
		\begin{align}\label{map 1}
			v=\frac{z-z_2}{z_1-z}\ ,
		\end{align}
		the operator $H_0$ can be simplified as
		\begin{align}\label{H_0 2}
			H_0=-\int_{\delta}^{\frac{1}{\delta}}\frac{dv}{2\pi}\ v T(v)+\cc\,,
		\end{align}
		where the Schwarzian derivative contribution vanishes since the mapping \eqref{map 1} is a M\"obius transformation, and the cut-off in \eqref{H_0 2} reads $\delta=\frac{z_1-z_2}{\epsilon}$.
		The recursion relation for the connected correlation functions of the stress tensor can be read out from the $TT$ OPE \cite{Belavin:1984vu}, given by
		\begin{align}
			\langle T(v_1)\cdots T(v_k)\rangle_{c}=\sum_{i=2}^{k}\left(\frac{2}{(v_1-v_i)^2}+\frac{\partial_{v_i}}{v_1-v_i}\right)\langle T(v_2)\cdots T(v_k)\rangle_{c}\,.
		\end{align}
		Using this recursion relation, it is straightforward to use \texttrademark\textit{Mathematica} to verify
		\begin{align}\label{eq: power of H0}
			& \langle (-2\pi H_0)^2 \rangle_c=\frac{cl}{3}\ ,\quad
			\langle (-2\pi H_0)^3 \rangle_c=-cl\ ,\nonumber\\
			& \langle (-2\pi H_0)^4 \rangle_c=4cl\ ,\quad
			\langle (-2\pi H_0)^5 \rangle_c=-20cl\ ,
		\end{align}		
		Inserting \eqref{eq: power of H0} into \eqref{eq: log gn} yields
		\begin{align}\label{gn result}
			\log g(n)=\frac{cl}{6}\left(-2n+n^2-n^3+n^4-n^5+\cdots\right)\,.
		\end{align}
		Evidently, this coincides with the usual result obtained in the twist field picture with replica index $n+1$, 
		\begin{align}\label{eq:C10}
			\langle \sigma_{n+1}(z_2)\tilde{\sigma}_{n+1}(z_1)\rangle=\frac{cl}{6}\left(\frac{1}{n+1}-n-1\right)=\frac{cl}{6}\left(-2n+n^2-n^3+n^4-\cdots\right)\ ,
		\end{align}
		where the last step expands for $|n|<1$. From comparing \eqref{gn result} and \eqref{eq:C10} we conjecture that $\langle (-2\pi H_0)^k \rangle_c$ fulfills a recursion relation as $\langle (2\pi H_0)^k \rangle_c=k\langle (2\pi H_0)^{k-1} \rangle_c$ starting from $\langle (2\pi H_0)^{2} \rangle_c=cl/3$. It is desirable to have an independent proof of this recursion relation.
	\end{appendix}
	\bibliographystyle{JHEP}
	\bibliography{../../library}
	
\end{document}